\documentclass[sigplan,screen,10pt, dvipsnames,svgnames,x11names,table]{acmart}
\renewcommand\footnotetextcopyrightpermission[1]{} 
\usepackage{glossaries}
\usepackage{xcolor}
\usepackage{xspace}
\usepackage{blindtext}
\usepackage{graphicx}
\usepackage{textcomp}
\usepackage{color, colortbl}
\usepackage{breakurl}
\usepackage{relsize}
\usepackage{cleveref}
\usepackage{hyperref}
\usepackage{subfloat}
\usepackage{multirow}
\usepackage{subfig}
\usepackage{setspace}
\usepackage{url}
\usepackage{listings}
\usepackage{color}
\usepackage{xcolor}
\usepackage{pifont}

\usepackage[margin=5pt,font={stretch=0.9}]{caption}

\def\BibTeX{{\rm B\kern-.05em{\sc i\kern-.025em b}\kern-.08em
		T\kern-.1667em\lower.7ex\hbox{E}\kern-.125emX}}

\definecolor{mymauve}{rgb}{0.58,0,0.82}

\hypersetup{
	colorlinks=True,
	linkcolor={green!50!black},
	citecolor={red!70!black},
	urlcolor={blue!70!black}
}

\newcommand{\sys}{\textsc{SecFL}\xspace}

\vspace{-1mm}
\title{\LARGE \bf
    \textsc{SecFL:}\xspace
    Confidential Federated Learning using TEEs 
}

\subtitle{\small Demo paper}
\author{Do Le Quoc}
\affiliation{\institution{Huawei Munich Research Center \country{Germany}}}
\author{Christof Fetzer}
\affiliation{\institution{TU Dresden \country{Germany}}}

\newacronym{ias}{IAS}{Intel Attestation Service}
\newacronym{sgx}{SGX}{Software Guard eXtensions}
\newacronym{tee}{TEE}{Trusted Execution Environment}
\newacronym{sdk}{SDK}{Software Development Kit}
\newacronym{cots}{COTS}{Components-off-the-Shelf}
\newacronym{os}{OS}{Operating System}
\newacronym{rop}{ROP}{Return-Oriented Programming}
\newacronym{afl}{AFL}{American Fuzzy Lop}
\newacronym{sigstruct}{SigStruct}{Enclave Signature Structure}
\newacronym{tcb}{TCB}{Trusted Computing Base}
\newacronym{secs}{SECS}{SGX Enclave Control Structure}
\newacronym{avx}{AVX}{Advanced Vector Extensions}
\newacronym{cet}{CET}{Control-flow Enforcement Technology}
\newacronym{einittoken}{EINITTOKEN}{EINIT Token}
\newacronym{flc}{FLC}{Flexible Launch Control}
\newacronym{faas}{FaaS}{Function as a Service}
\newacronym{iaas}{IaaS}{Infrastructure as a Service}
\newacronym{saas}{SaaS}{Software as a Service}
\newacronym{sfi}{SFI}{Software Fault Isolation}
\newacronym{cas}{CAS}{Configuration and Attestation Service}

\begin{document}
\begin{abstract}
Federated Learning (FL) is an emerging machine learning paradigm that enables multiple clients to jointly train a model to take benefits from diverse datasets from the clients without sharing their local training datasets.  FL helps reduce data privacy risks. Unfortunately, FL still exist several issues regarding privacy and security. First, it is possible to leak sensitive information from the shared training parameters. Second, malicious clients can collude with each other to steal data, models from regular clients or corrupt the global training model. 
To tackle these challenges, we propose \sys - a confidential federated learning framework that leverages Trusted Execution Environments (TEEs).  \sys performs the global and local training inside TEE enclaves to ensure the confidentiality and integrity of the computations against powerful adversaries with privileged access. \sys provides a transparent remote attestation mechanism, relying on the remote attestation provided by TEEs, to allow clients to attest the global training computation as well as the local training computation of each other. Thus, all malicious clients can be detected using the remote attestation mechanisms. 
 
\end{abstract}

\maketitle
\thispagestyle{empty}
\pagestyle{empty}

\section{Introduction}
\label{sec:intro}

FL~\cite{fl} is an emerging machine learning technique which allows participating clients to collaboratively train a joint global machine learning model without sharing their local training data. FL reduces privacy risks for the local training data which may be highly sensitive relating to personal finances, political views, health, etc. Thus, it has been used widely in the industry since it helps companies comply with regulations on issues regarding the way in which personal data is handled and processed - such as EU's General Data Protection Regulation (GDPR)~\cite{gdpr}. 
The core idea of FL is that each client trains a local model, rather than sharing training data to a centralized training system which is deployed in an untrusted environment, e.g., a public cloud. 
For each iteration, the clients send their local training parameters to the central system to train a global model which takes benefits from all local training data from clients. Typically, the central system aggregates the local training parameters from the clients and sends the aggregated parameters back to them. This training process is repeated until it converges or the global model reaches a certain desired accuracy.  An example of FL in real-life deployment is that several hospitals collaborate to develop a shared machine learning model based on their patient data to detect a disease at an early stage.  Each hospital trains its data locally, shares the local model with the central training system, and receives the global model in each iteration. 

While promising at first glance, FL paradigm suffers several vulnerabilities. First, an attacker with privileged/root accesses can easily obtain the training models (\raisebox{-1pt}{\ding{202}}). The attacker can also compromise the privacy of individuals in the training data by inferring it from parameters of the global model~\cite{deeplearning-DP}. 
Therefore, the training models need to be protected at rest, in transit, and in use. 
Second, a large number of malicious clients may collude with each other to reveal local data and local models of the remaining clients~\cite{mugunthan2020privacyfl} (\raisebox{-1pt}{\ding{203}}).
Last but not least, these malicious clients can tamper their local training data or parameters updates forwarded to the central training system to corrupt the global model~\cite{xie2020fall,fang2020local} (\raisebox{-1pt}{\ding{204}}).

To handle the issues \raisebox{-1pt}{\ding{202}} \raisebox{-1pt}{\ding{203}}, state-of-the-art solutions rely on a privacy-preserving mechanism such as differential privacy or secure multiparty computation (MPC). The disadvantage of the differential privacy mechanism is that it reduces the performance of the global training model regarding utility or accuracy. Meanwhile, the solutions based on secure multiparty computation incur significant overhead~\cite{securetf,tensorscone}.  To cope with issue \raisebox{-1pt}{\ding{204}},  several Byzantine-robust federated learning mechanisms have been proposed~\cite{blanchard2017machine,xie2020fall,fang2020local, bhagoji2019analyzing}. The core idea behind these mechanisms is to reduce the impact of statistical outliers during model updates in the federated learning system. 
However, recent works~\cite{xie2020fall,fang2020local,bhagoji2019analyzing}  show that the mitigation of the impact is still not enough to protect the utility of the global model. The malicious clients can still affect the accuracy of the global model trained by a Byzantine-robust mechanism by carefully tampering with their model parameters sent to the central training system~\cite{cao2021provably}.

In this work, we overcome these limitations by building a confidential federated learning system called \sys using TEEs, e.g., Intel SGX.
\gls{tee} technologies, such as Intel \gls{sgx}, have gained much attention in the industry~\cite{AzureSGX, IBMCloudSGX,singh2021enclaves} as well as in academia~\cite{scone,costan2016intel,tsai2017graphene,sgx-pyspark,ozga2021perun,perun2,securetf,singh2020enclaves,teemon,TSR,avocado,weles}.
To ensure confidentiality and integrity of applications, \glspl{tee} execute their code and data inside an encrypted memory region called \emph{enclave}. 
Adversaries with privileged access cannot read or interfere with the memory region and only the processor can decrypt and execute the application inside an enclave.  
In addition, TEEs such as Intel SGX also provide a mechanism for users to verify that the \gls{tee} is genuine and that an adversary did not alter their application running inside TEE enclaves. The verification process is called \emph{Remote Attestation}~\cite{costan2016intel} and allows users to establish trust in their application running inside an enclave on a remote host.

We leverage \glspl{tee} to handle issue \raisebox{-1pt}{\ding{202}}, by providing end-to-end encryption in \sys. \sys encrypts input training data and code (e.g., Python code) and performs  all training computations including local training and global training insides TEE enclaves. \sys enables all model updates via TLS connections between the enclave of clients and the enclaves of the central training computation.  Thus attackers with privileged accesses cannot violate the integrity and confidentiality of the input training data, code, and models. \sys also ensures the {\em freshness} of the input training data, models, and, by applying an advanced asynchronous monotonic counter service~\cite{adam-cs}.  
We tackle issues \raisebox{-1pt}{\ding{203}} \raisebox{-1pt}{\ding{204}} by developing in \sys a  {\em Security Policy Manager} component based on the remote attestation mechanism supported by TEEs~\cite{palaemon,intel-remote-attestation}. The component ensures the integrity of input data and training code, i.e., it makes sure that training computations are running with correct code, correct input data and not modified by anyone, e.g., an attacker or malicious client. This component also monitors and attests to the compliance of participated clients with the pre-defined agreement before collaborating to train the global machine learning model. In addition, \sys can clone the global training computation and randomly take a sample of clients for the training computation. This helps to detect outliers regarding the utility which helps to solve  issue  \raisebox{-1pt}{\ding{204}}.
Our preliminary evaluation shows that \sys can ensure the confidentiality and integrity of federated learning computations while maintaining the same utility/accuracy of the training computations. 

\section{Confidential Federated Learning}
\vspace{-1mm}

\begin{figure}[t]
	\centering
	\includegraphics[scale=0.3]{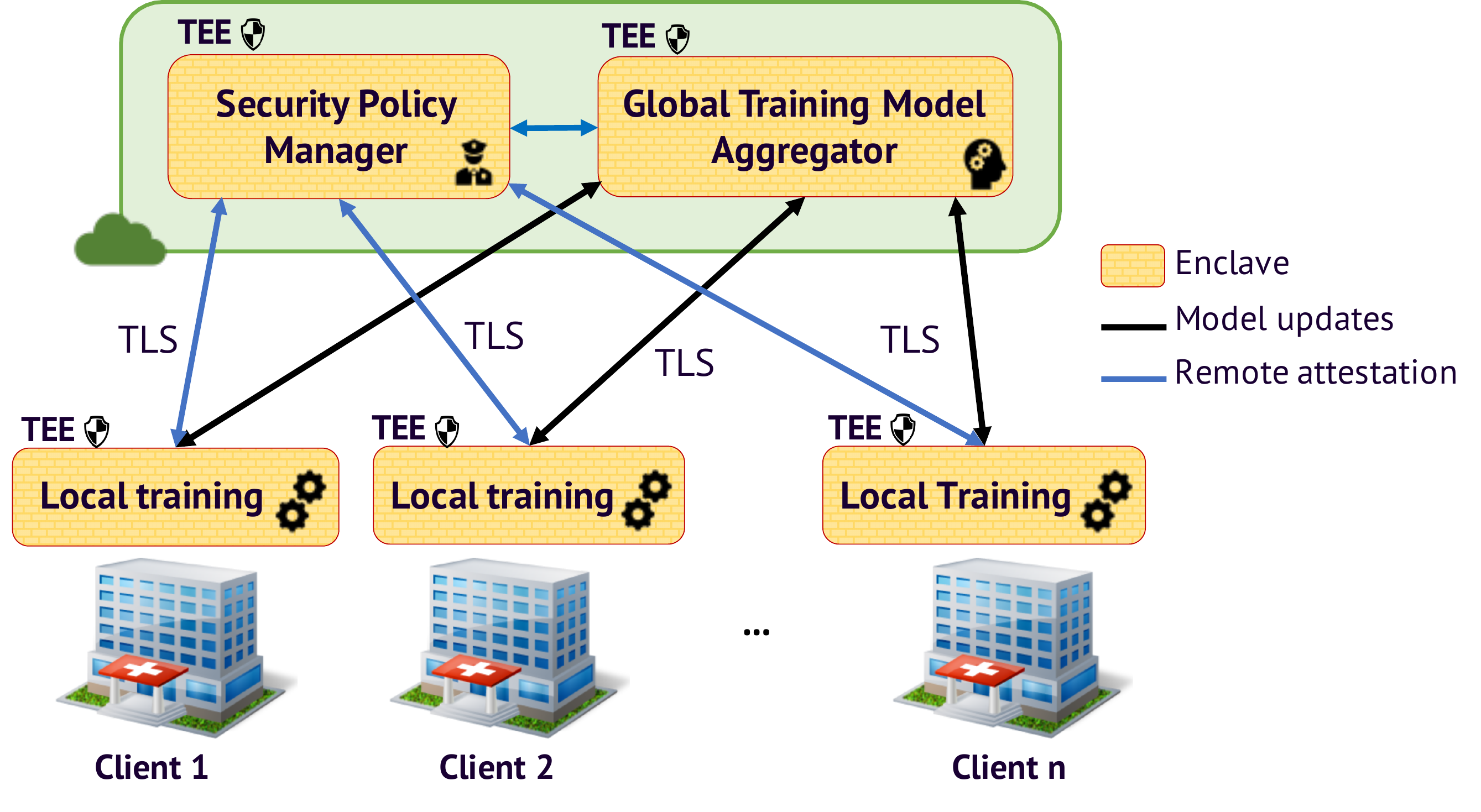}
	\caption{\sys Architecture for Federated Learning.}
	\label{fig:secfl}
\end{figure}

Figure~\ref{fig:secfl} shows the architecture of \sys.  
The main goal of \sys  is not only to ensure the confidentiality, integrity and freshness of input data, code, and machine learning models but also to enable multiple clients (who do not necessarily trust each other) to get the benefits of collaborative training without revealing their local training data.  
In \sys, each client performs the local training also inside TEE  enclaves to make sure that no one tamper input data or training code during the computations. To govern and coordinate the collaborative machine learning training computation between clients, we design in \sys a trusted management component, called {\em Security Policy Manager} which maintains  security policies based on the agreement among all clients to define the access control over global training computation, the global training model, also the code and input data used for local training at each client. Security Policy Manager automatically and transparently performs remote attestation to make sure the local computations are running correct code, correct input data, and on correct platforms as the agreement. It only allows clients to participate in the global training after successfully performing the remote attestation process. 
It also conducts the remote attestation on the enclaves that execute the global training in a cloud, to ensure that no one at the cloud provider side modify the global training aggregation computation. \sys encrypts the training code, and Security Policy Manager only provides the key to decrypt it inside enclaves after the remote attestation.  Secrets including keys for encryption/decryption in each policy are generated by the Security Policy Manager also running inside Intel SGX enclaves and cannot be seen by any human or client.  Examples of the policies can be found in ~\cite{sconedocs,palaemon}.

After receiving the agreed security policies from clients, Security Policy Manager strictly enforces them. It only passes secrets and configuration to  applications (i.e., training computations), after attesting them. The training computations are executed inside Intel SGX enclaves and associated with policies provided and pre-agreed by clients. The training computations are identified by a secure hash and the content of the files (input data) they can access. Secrets can be passed to applications as command-line arguments, environment variables, or can be injected into files. The files can contain environment variables referring to the names of secrets defined in the security policy. The variables are transparently replaced by the value of the secret when an application that is permitted to access the secrets reads the file. 
We design the component  Security Policy Manager in the way that we can delegate the management of it to an untrusted party, e.g., a cloud provider, while clients can still trust that their security policies for protecting their properties are safely maintained and well protected. 
In \sys, clients can attest Security Policy Manager component, i.e., they can verify that it runs the expected unmodified code, in a correct platform before uploading security policies.
We implement \sys using Intel OpenFL~\cite{intel-openfl} --- a distributed federated machine learning framework 
We run the local and global training computations inside SGX enclaves using SCONE~\cite{scone} --- a shielded execution framework to enable unmodified applications to run inside SGX enclaves. In the SCONE platform, the source code of an application is recompiled against a modified standard C library (SCONE libc) to facilitate the execution of system calls. The address space of the application stays within an enclave. 
In \sys, the input training data and code are encrypted using the file system shield of SCONE, and then decrypted and processed inside SGX enclaves which cannot be accessed even by strong attackers with root access. We rely on our previous works~\cite{palaemon,sconedocs} to implement Security Policy Manager. A demo of \sys is publicly available in~\cite{secfl-demo}.

\bibliographystyle{ACM-Reference-Format}
\bibliography{main}

\end{document}